\documentclass[12pt,preprint]{aastex}
\usepackage{psfig,lscape}
\def\lap{\lower.5ex\hbox{$\; \buildrel < \over \sim \;$}}
\def\gap{\lower.5ex\hbox{$\; \buildrel > \over \sim \;$}}

\def\ergcm2s{${\rm erg\ cm^{-2}\ s^{-1}}$}

\def\ergscm2s{${\rm erg\ cm^{-2}\  s^{-1}}$}

\def\cm-2{${\rm cm^{-2}}$}

\begin{document}

\title{Optical Constraints on an X-ray Transient Source in M31}

\author{Benjamin F. Williams\altaffilmark{1}, Michael
R. Garcia\altaffilmark{1}, Frank A. Primini\altaffilmark{1}, Jeffrey
E. McClintock\altaffilmark{1}, and Stephen S. Murray\altaffilmark{1}}
\altaffiltext{1}{Harvard-Smithsonian Center for Astrophysics, 60
Garden Street, Cambridge, MA 02138; williams@head.cfa.harvard.edu;
garcia@head.cfa.harvard.edu; fap@head.cfa.harvard.edu; jem@head.cfa.harvard.edu;
ssm@head.cfa.harvard.edu}


\begin{abstract}

We have detected a transient X-ray source in the M31 bulge through a
continuing monitoring campaign with the {\it Chandra} ACIS-I camera.
The source was detected at R.A.=00:42:33.428 $\pm$ 0.11$''$,
Dec.=+41:17:03.37 $\pm$ 0.11$''$ in only a single observation taken
2004-May-23.  Fortuitous optical $HST$/ACS imaging of the transient
location prior to the X-ray outburst, along with follow-up $HST$/ACS
imaging during and after the outburst, reveals no transient optical
source brighter than $B$ (equivalent) = 25.5.  The location of the
source and its X-ray properties suggest it is a low mass X-ray binary
(LMXB).  Assuming the transient is similar to many Galactic X-ray
novae, the X-ray luminosity of (3.9$\pm$0.5)$\times$10$^{37}$ erg
s$^{-1}$ and the upper-limit on the optical luminosity provide a
prediction of $<$1.6 days for the orbital period of the binary system.

\end{abstract}

\section{Introduction}

Many low-mass X-ray binaries (LMXBs) that undergo bright, transient
X-ray bursts have been shown to harbor compact objects with masses
$>$3~M$_{\odot}$ (see \citealp{mcclintock2004}, and references
therein).  These sources are some of the most securely identified
black holes known.  Such sources are therefore of great interest for
future studies of black hole accretion disk physics and general
relativity.

The {\it Chandra} X-ray observatory is well-suited to searching for
similar X-ray sources in nearby galaxies.  In particular, the bulge of
M31 can be entirely searched for bright, transient X-ray events with a
single 5 ks observation.  Monitoring of M31 by {\it Chandra} has shown
that such transient sources appear in the M31 bulge about once each
month \citep{williams2004}.  These efforts have been successful in
finding dozens of transient sources
\citep{kong2002,distefano2004,williams2004}.

Recently, this monitoring effort has been combined with follow-up
$HST$ observations, and optical counterparts for some transient X-ray
sources have been found, placing new constraints on the physical
properties of these potential black hole binary systems
\citep{williams2004,williams2005bh1,williams2005bh2}.  With a
sufficient number of optical counterparts, the orbital period
distribution of these likely black hole binaries can be determined.
This distribution is a fundamental observable parameter that must be
matched by any model of binary stellar evolution.

Here, we report a bright outburst from an X-ray source in M31.  $HST$
imaging prior to, during, and after the outburst reveals no optical
sources that exhibited strong variability during the X-ray outburst
within the location uncertainty, suggesting that the optical
counterpart of the X-ray nova (XRN) was fainter than $B=25.5$.
Section 2 discusses the details of the data, including the reduction
and analysis techniques used.  Section 3 provides the results of the
analysis, and \S~4 explains the implications of these results.
Finally, \S~5 summarizes our conclusions.

\section{Data Reduction and Analysis}

\subsection{X-ray}

We obtained observations of the M31 bulge with the {\it Chandra}
ACIS-I camera on 2004-January-31 (ObsID~4681), 2004-May-23
(ObsID~4682), and 2004-July-17 (ObsID~4719).  Observations 4681 and
4682 were Guaranteed Time Observations supplied by S. Murray.  All of
the data were obtained in ``alternating readout mode'' which reduces
event pileup for bright sources but lowers the effective exposure by
$\sim20\%$.  Observation 4681 was observed centered on
R.A.=00:42:44.4, Dec.=41:16:08.3 with a roll angle of 305.6 degrees
for an effective exposure time of 4.09 ks.  Observation 4682 was
centered on the same coordinates with a roll angle of 80.0 degrees for
an effective exposure time of 3.95 ks, and observation 4719 was
observed centered on R.A.=00:42:44.3, Dec.=41:16:08.4 with a roll
angle of 116.8 degrees for an effective exposure time of 4.12 ks.

These observations were all reduced using the software package CIAO
v3.1 with the CALDB v2.28.  We created exposure maps for the images
using the task {\it
merge\_all},\footnote{http://cxc.harvard.edu/ciao/download/scripts/merge\_all.tar}
and we found and measured positions and fluxes of the sources in the
image using the CIAO task {\it
wavdetect}.\footnote{http://cxc.harvard.edu/ciao3.0/download/doc/detect\_html\_manual/Manual.html}
Each data set detected sources down to (0.3--10 keV) fluxes of
$\sim$6$\times$10$^{-6}$ photons cm$^{-2}$ s$^{-1}$ or 0.3--10 keV
luminosities of $\sim$10$^{36}$ erg s$^{-1}$ for a typical X-ray
binary system in M31.

We aligned the coordinate system of the X-ray images with the optical
images of the Local Group Survey (LGS; \citealp{massey2001}).  These
images have an assigned J2000 (FK5) world coordinate system accurate
to $\sim$0.25$''$, and they provided the standard coordinate system to
which we aligned all of our data for this project.  The positions of
X-ray sources with known globular cluster counterparts were aligned
with the globular cluster centers in the LGS images using the
IRAF\footnote{IRAF is distributed by the National Optical Astronomy
Observatory, which is operated by the Association of Universities for
Research in Astronomy, Inc., under cooperative agreement with the
National Science Foundation.} task {\it ccmap}, allowing adjustments
of the pixel scale as well as rotation and shifts in $X$ and $Y$.  The
alignment had errors of 0.073$''$ in R.A. and 0.034$''$ in Dec. as
determined by {\it ccmap}.

In addition, we measured the position errors for the X-ray sources
using the IRAF task {\it imcentroid}, which projects the counts from
the source onto each axis and calculates the error in the position by
dividing the standard deviation of the pixel positions of all of the
source counts by the square root of the number of counts.  Because the
pixels in {\it Chandra} images are aligned with north up and east to
the left, the X position error was taken to be the R.A. error, and the
Y position error was taken to be the Dec. error.

We cross-correlated the X-ray source positions against all previously
published X-ray catalogs and the {\it
Simbad}\footnote{http://simbad.u-strasbg.fr/} database to look for any
new, bright X-ray source likely to be an X-ray nova (XRN).  Herein we
focus on one bright source in particular which was detected only in
observation 4682 at R.A.=00:42:33.428, Dec.=+41:17:03.37.  This source
was detected once by ROSAT in 1990 ([PFJ93]~31;
\citealp{primini1993}), but it has not been detected since.  We name
this source CXOM31~J004233.4+411703, following the naming convention
described in \citet{kong2002}.  We also give this source the short
name, r2-71, based on the source position in M31 using the description
given in \cite{williams2004}.  The source is 2.0$'$ west and 0.9$'$
north of the M31 nucleus.  Upper-limits to the X-ray flux at this
position in observations 4681 and 4719 were measured by determining
the flux necessary to produce a detection 4-$\sigma$ above the
background flux.

We extracted the X-ray spectrum of r2-71 from the detection in
observation 4682, which contained 198 counts, using the CIAO task {\it
psextract}.\footnote{http://cxc.harvard.edu/ciao/ahelp/psextract.html}
We binned the spectrum in energy so that each bin contained $\gap$10
counts to allow for standard $\chi^2$ statistics.  We then fit the
spectrum using CIAO 3.1/Sherpa.  Two spectral models were fitted to
the spectrum: a power-law with absorption and a disk blackbody model
with absorption.  The spectrum did not contain sufficient information
to provide useful constraints on the foreground absorption.  We
therefore fixed the absorption to the typical Galactic foreground
value (6$\times$10$^{20}$ cm$^{-2}$).  This value provided acceptable
fits for both model types.  Results are discussed in \S~3.

\subsection{Optical}

We obtained three sets of {\it HST} ACS data.  The first ACS image was
taken 2004-January-23 in pursuit of another transient source.  This
observation was pointed at R.A.=00:42:43.86, Dec.=41:16:30.1 with an
orientation of 55 deg, fortuitously containing the location of r2-71
in the northern corner of the image even though r2-71 had not been
detected in any {\it Chandra} observations.  The second observation
was taken 2004-June-14.  Intended to search for the optical
counterpart of r2-71, this observation was pointed at
R.A.=00:42:33.39, Dec.=41:17:43.4 with an orientation of 260~deg. The
slightly northern pointing was done to avoid a dangerously bright star
from landing on the CCD.  The third observation was taken on
2004-August-15, after the X-ray source had faded.  This observation
was pointed at R.A.=00:42:41.5, Dec.=41:17:00.0 with an orientation of
220 deg in order to pursue another transient source in the field.

All three observations were taken using the standard ACS box 4-point
dither pattern to recover the highest possible spatial resolution.
All exposures were taken through the F435W filter.  The total exposure
times were 2200 seconds for each data set.  We aligned and drizzled
each set of 4 images into high-resolution (0.025$''$ pixel$^{-1}$)
images using the PyRAF\footnote{PyRAF is a product of the Space
Telescope Science Institute, which is operated by AURA for NASA.} task
{\it multidrizzle},\footnote{multidrizzle is a product of the Space
Telescope Science Institute, which is operated by AURA for
NASA. http://stsdas.stsci.edu/pydrizzle/multidrizzle} which has been
optimized to process ACS imaging data.  The task removes cosmic ray
events and geometric distortions, and it combines the dithered frames
together into one final photometric image.

\begin{figure}
\centerline{\psfig{file=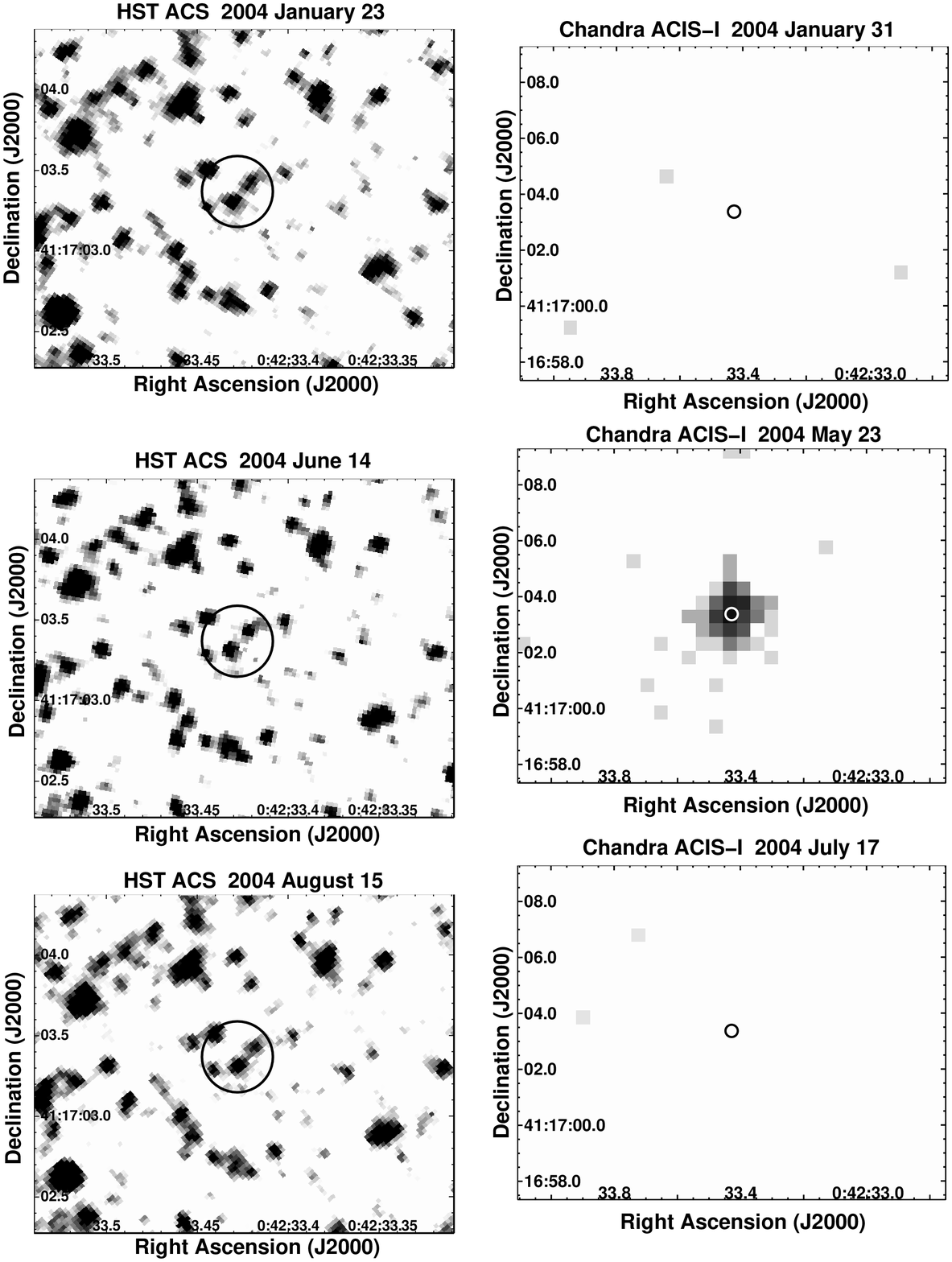,height=6.0in,angle=0}}
\caption{{\it Left panels: HST} ACS images of the location of r2-71
taken before, during, and after the outburst.  The black circles mark
the 2$\sigma$ error for the X-ray position of r2-71. No significant
variability was identified inside the error circle. {\it Right panels:
Chandra} ACIS-I images of r2-71 before, during, and after the
outburst.  The error circles in the left panels correspond to the
circles on the X-ray images.}
\label{ims}
\end{figure}

We aligned the $HST$ images to the LGS coordinate system with {\it
ccmap} using stars common to both data sets.  The resulting alignment
had rms errors of 0.04$''$ (less than 1 ACS pixel).  The consistency
of this alignment can be seen by the excellent agreement between the
resulting coordinate systems of the three $HST$ images, each
independently aligned with the LGS coordinate system, shown in
Figure~\ref{ims}.

We processed the relevant sections of the final images with DAOPHOT~II
and ALLSTAR \citep{stetson} to obtain photometry for the resolved
stars within the error circle of the X-ray transient.  This region is
extremely crowded with stars at only 2.2$'$ from the M31 nucleus.  The
faintest source visible in the error circles in the images shown in
Figure~\ref{ims} is $B=26.7$; the brightest is $B=24.7$ (on the
northeast rim of the circle).

\begin{figure}
\centerline{\psfig{file=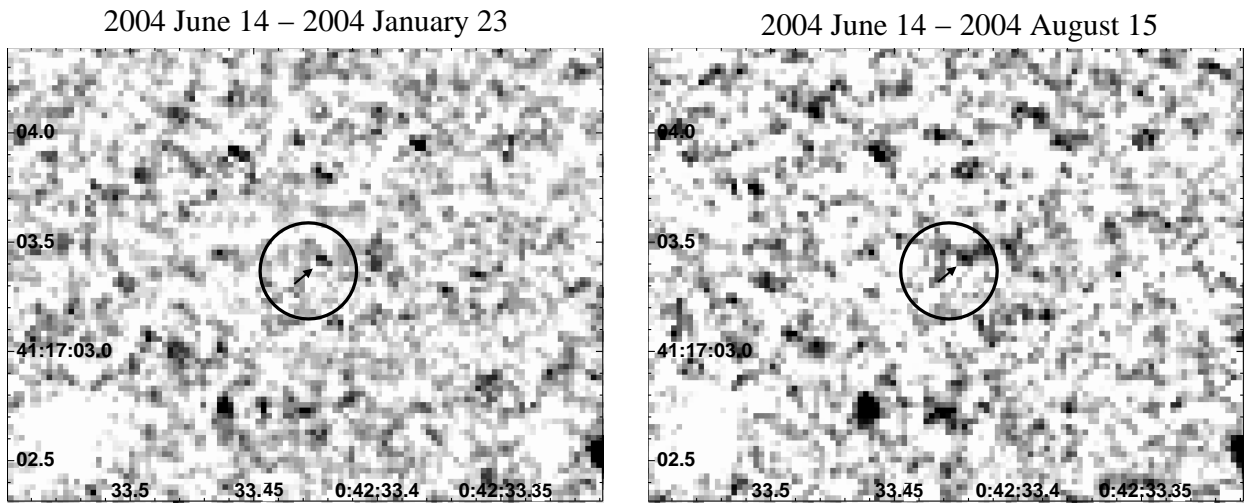,width=6.5in,angle=0}}
\caption{{\it Left panel:} The 2004-June-14 ACS image with the
2004-May-23 ACS image subtracted away.  The greyscale is set so that
darker areas had more counts in the 2004-June-14 data.  The black
circle shows the X-ray position error for r2-71.  The arrow marks the
position of greatest difference within the error circle. {\it Right
panel:} The 2004-June-14 ACS image with the 2004-August-15 ACS image
subtracted away.  The greyscale and symbols are the same as in the
left panel.}
\label{diff}
\end{figure}

We subtracted the images taken before and after the X-ray nova from
the 2004-June-14 observation, during which the X-ray source was likely
to be active.  Prior to subtraction, the images were transformed to
have pixels aligned in North-up, East-left orientation.  These
transformations were performed with the IRAF tasks {\it geomap} and
{\it geotran}.  The relevant sections of the subtracted images are
shown in Figure~\ref{diff}.

\section{Results}

\subsection{X-ray}

The three X-ray images from our $Chandra$ observations are shown in
Figure~\ref{ims}.  These images clearly show r2-71 detected only in
the 2004-May-23 observation.  This detection had a 0.3--10 keV flux of
(1.7$\pm$0.1$)\times$10$^{-4}$ photons cm$^{-2}$ s$^{-1}$.  The
previous observation and following observation had 4$\sigma$ upper
limits of $<$1.4$\times$10$^{-5}$ photons cm$^{-2}$ s$^{-1}$.  The
second upper limit shows that the source decayed by a factor of at
least 12 in 55 days.  Therefore the $e$-folding decay time of r2-71
was $\leq$22 days.

The non-detection of this source in the survey of \cite{kong2002}
provides a 0.3--7 keV flux upper limit of $<$8$\times$10$^{-7}$
photons cm$^{-2}$ s$^{-1}$, showing that the source changed in flux by
more than a factor of 100.  Therefore r2-71 is certainly a transient
X-ray source in M31.

The errors in the centroid determination of the X-ray source were
0.08$''$ in R.A. and 0.10$''$ in Declination.  We added these errors
in quadrature to the errors in the alignment of the X-ray and optical
images (0.073$''$ in R.A. and 0.034$''$ in Declination) to obtain the
final (1$\sigma$) position errors of 0.11$''$ and 0.11$''$ in R.A. and
Declination respectively.  These errors resulted in the 2$\sigma$
error circle shown in Figure~\ref{ims}.

The X-ray spectrum of r2-71 was well-fitted by both the power-law and
the disk-blackbody models.  Fortunately, both of these fits give the
same measurement for the absorption-corrected 0.3--7 keV flux and the
corresponding 0.3--7 keV luminosity.  The best-fitting power-law has a
slope of 1.5$\pm$0.1 with $\chi^2/\nu$=17.87/17 (probability = 0.40).
This fit yields an unabsorbed 0.3--7 keV flux of
($5.4\pm0.5$)$\times$10$^{-13}$ erg cm$^{-2}$ s$^{-1}$. The best
fitting disk blackbody has an inner disk temperature of kT=1.4$\pm$0.2
keV, an inner disk radius of $(6\pm2)/cos^{1/2}(i)$ km with a
$\chi^2/\nu=16.38/17$ (probability=0.50).  The resulting unabsorbed
0.3--7 keV flux is ($4.9\pm2.2$)$\times$10$^{-13}$ erg cm$^{-2}$
s$^{-1}$.

Assuming a distance to M31 of 780 kpc \citep{williams2003}, the
results are both consistent with the X-ray luminosity of
(3.9$\pm$0.4)$\times$10$^{37}$ erg s$^{-1}$ obtained from the
power-law fit, similar to the luminosity seen in the 1990 ROSAT data
by \cite{primini1993}.  The spectrum of r2-71 is in the normal range
of M31 X-ray transients as measured by Williams et al. (in
preparation), with hardness ratios of $(M-S)/(M+S) = 0.59\pm0.11$ and
$(H-S)/(H+S) = 0.52\pm0.11$, where S, M, and H represent the number of
counts detected in the energy ranges 0.3--1.0, 1.0--2.0, and 2.0--7.0
keV, respectively.  Gaussian errors were measured for the
background-subtracted number of counts in each energy bin with the
CIAO task {\it dmextract}.

\subsection{Optical}

Analysis of the optical data initially provided a few possible
variable stars within the r2-71 error circle.  We scrutinized each
possibility through aperture photometry, completeness tests, and
difference imaging.  The results show no strong detection of optical
variability and provide an upper-limit to the $B$ magnitude of any
highly variable counterpart to r2-71.

The region of interest for all three of our optical observations is
shown in the images in Figure~\ref{ims}.  The DAOPHOT II output
revealed one star in the error circle that was significantly brighter
in the 2004-June-14 observation, when the X-ray source was most likely
active.  The star at R.A.=00:42:33.414, decl.=41:17:03.47, was
measured to have $B=25.50\pm0.06$ in the 2004-June-14 observation.
DAOPHOT II failed to find this star in the 2004-January-23
observation.  Because inspection of the 2004-January-23 image reveals
a source at this location, this non-detection was likely due to the
effects of the bright neighboring star 0.05$''$ to the southeast.
Aperture photometry of the location in the 2004-January-23 observation
with a 0.075$''$ radius aperture yields $B=26.0\pm0.1$.  DAOPHOT II
measured the star to be $B=25.98\pm0.09$ in the 2004-August-15
observation. Therefore according to the standard errors an increase in
brightness during the 2004-June-14 at the 4$\sigma$ confidence level
occurred; however, the standard errors do not take into account the
uncertainty introduced by the close brighter neighbor.  Any added
uncertainty due to crowding would decrease the significance of this
brightness increase.

In addition to this suspicious counterpart candidate, there were 2
fainter stars detected by DAOPHOT II in the error circle of the
2004-June-14 observation that were not detected in the other 2
observations, when the X-ray source was not active.  These stars had
$B$ magnitudes of 26.6$\pm$0.2 and 26.7$\pm$0.2 in the 2004-June-14
observation.  These stars may not have been detected in the other
observations because crowding issues caused our photometry to be
incomplete at these faint magnitudes.

\begin{figure}
\centerline{\psfig{file=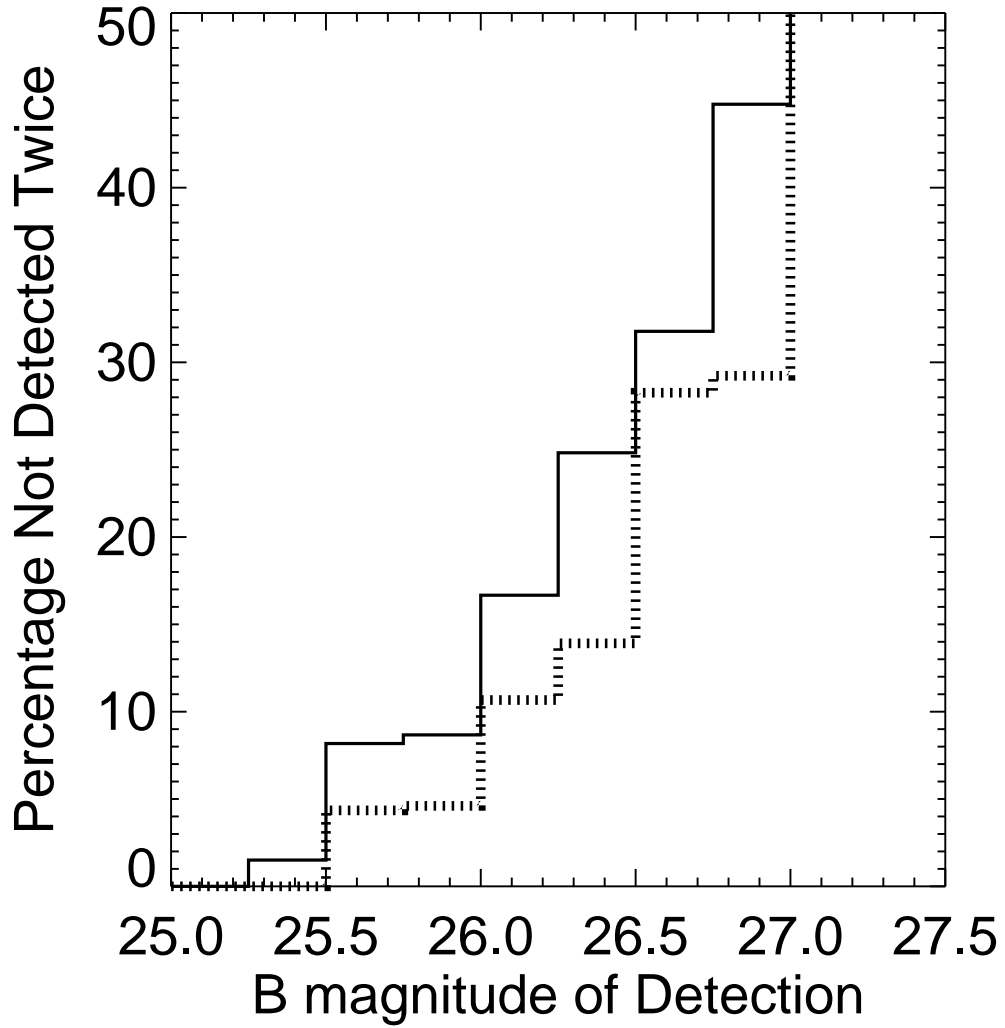,height=6.0in,angle=0}}
\caption{{\it Solid Histogram:} the percentage of stars within 3$''$
of r2-71 detected by DAOPHOT II in the 2004-June-14 ACS image that
were not detected in by DAOPHOT II in the 2004-August-15 ACS image as
a function of $B$ magnitude in the 2004-June-14 ACS image.  {\it
Dotted Histogram:} the percentage of stars within 3$''$ of r2-71
detected by DAOPHOT II in the 2004-August-15 ACS image that were not
detected in by DAOPHOT II in the 2004-June-14 ACS image as a function
of $B$ magnitude in the 2004-August-15 ACS image.  The results suggest
that the completeness of the data begins to decrease at $B=25.5$.}
\label{comp}
\end{figure}

We determined the completeness of the optical data in the area of
r2-71 by comparing the DAOPHOT II output from the 2004-June-14
observation to those of the 2004-August-15 observation.
Figure~\ref{comp} shows two histograms.  The solid histogram shows the
percentage of stars detected by the DAOPHOT analysis in the
2004-June-14 observation within 3$''$ of the center of the error
circle but not detected by the same analysis in the 2004-August-15
observation.  The dotted histogram shows the percentage of stars
detected by the DAOPHOT analysis in the 2004-August-15 observation
within 3$''$ of the center of the error circle but not detected by the
same analysis in the 2004-June-14 observation. The number of lost
stars begins to increase at $B=25.5$, suggesting that the completeness
of the data begins to decrease at that magnitude, most likely due to
the crowding in this dense region of M31.  This result is consistent
with the failure of DAOPHOT II to detect the $B=26.0$ star in the
error circle in the 2004-January-23 observation, even though there
appears to be emission at that location in the image in
Figure~\ref{ims}. Therefore all of the variable candidates in the
r2-71 error circle are attributable to crowding and completeness
issues, as they are all fainter than $B=25.5$.

The lack of any strong variability detection within the r2-71 error
circle is confirmed with the difference images shown in
Figure~\ref{diff}.  The most variable location in the difference
images is marked with arrows; this variability is not statistically
significant.  The DAOPHOT analysis did not measure a brightness
increase for this star in the 2004-June-14 data.  As a second check
for a possible brightness increase, we performed aperture photometry
of the location in all three observations with a 0.1$''$ radius
aperture.  The location had $B=25.07\pm0.07$ in the 2004-June-14
observation and $B=25.23\pm0.08$ in the other observations ($\Delta B
= 0.16\pm0.11$), showing variability of only 1.5$\sigma$.  We
discounted this low-significance peak in the difference image as a
counterpart candidate for r2-71.

Concisely, no variable star inside the error circle of r2-71 was found
other than suspicious DAOPHOT detections affected by crowding and
completeness.  Therefore, we were unable to identify the optical
counterpart to r2-71; however, our completeness results suggest that
any highly variable counterpart must have had $B\geq25.5$ during the
2004-June-14 $HST$ observation.

\section{Discussion}\label{discussion}

\subsection{Duty Cycle}

Our search of the literature found one previous detection of r2-71 in
1990.  The detection was in only one ROSAT observation (source 31 in
\citealp{primini1993}), and the lack of detections of outbursts of
this source before or since
\citep{trinchieri1991,kong2002,williams2004} helps to constrain the
duty cycle of this system.  M31 was observed by {\it Einstein} in the
summers and winters of 1979 and 1980 \citep{trinchieri1991} and by
ROSAT in the summers of 1990 \citep{primini1993}, 1991
\citep{supper1997}, 1992 \citep{supper2001}, 1994, and 1995.  We
searched the ROSAT {\it HRI} images of the M31 bulge taken in the
summers of 1994 and 1995 and found no detections of r2-71.  It was
therefore detected only once in about 9 months of monitoring spanning
7 years before {\it Chandra}.  These observations constrain the duty
cycle of the source to be $\lap$0.1, since a larger duty cycle would
have allowed more than one detection in these early data sets.

Now, in 2004, r2-71 has been seen for the first time in about 5 years
of {\it Chandra} monitoring, going back to late 1999
\citep{williams2004,kong2002}.  Assuming the 1990 outburst also lasted
$\sim$1 month, the source has been active for at least 2 months of a
14 year timespan, providing a lower limit on its duty cycle of
$>$0.01. Adding the $\sim$8 months per year for 5 years of {\it
Chandra} monitoring to the 9 months of monitoring before {\it Chandra}
implies 2 months of activity in $\sim$49 months of monitoring, or a
duty cycle of $\sim$0.04.  Therefore, our best estimate of the duty
cycle of r2-71 is $\sim$0.04, and it can be reliably constrained to
the range 0.01--0.1.

\subsection{Orbital Period}

The {\it HST} data provide indirect evidence that r2-71 is an LMXB.
This preliminary classification allows us to predict the range in
which the orbital period of the system will fall assuming the system
is similar to Galactic LMXB transient systems.

The $HST$ data set rules against the possibility that the X-ray source
is an HMXB.  Even the brightest star in the error circle of r2-71 has
$B=24.7$, which implies M$_B = -0.2$ (assuming $m-M = 24.47$ and $A_B
= 0.4$).  Even so, this $B$-band luminosity is fainter than massive O
and B type stars. Furthermore, this brightest star is not considered a
counterpart candidate because it did not show significant variability.
Because r2-71 is not an HMXB, we continue under the assumption that it
is an LMXB.

Van Paradijs \& McClintock (1994) identified an empirical correlation
between the optical/X-ray luminosity ratios of LMXBs in outburst and
their orbital periods. Their model assumes that the optical emission
arises from X-ray heating of the accretion disk.  Larger disks form in
longer period systems and glow brighter in the optical than smaller
disks.  In this model, the faint upper-limit on the optical brightness
of r2-71 would suggest that it is a small accretion disk system with a
short orbital period.  Counter-examples to the correlation exist, like
XTE~J1118+480 \citep{williams2005bh1}; however, such counter-examples
usually stand out as odd in other ways.  For example, XTE~J1118+480
was fainter and harder than typical XRNe.  Therefore, for the purposes
of the prediction, we assume that r2-71 is an XRN similar to the many
Galactic LMXB transient events that fit the correlation well.  A few
of these are described in detail below.

We checked the applicability of the correlation to the specific case
of our data because our X-ray and optical data are not precisely
contemporaneous.  If the errors in the correlation and optical
luminosity are taken into account, our investigations show that the
correlation provides reliable period range predictions for both
``classical'' and more recently discovered Galactic XRNe, even if the
optical luminosity is measured 3 weeks after the X-ray luminosity.

First, we checked the application of the correlation to ``classical''
Galactic XRNe, those with smooth exponential decays (e.g. A0620-00,
Nova Mus, GRO~0422+32, etc.). These types of events have optical decay
timescales that average $\sim$2.2 times longer than their X-ray decay
timescales \citep{chen1997}. For example, A0620-00 has an $e$-folding
optical decay time of $\sim$75 days and an X-ray decay time of
$\sim$25 days \citep{esin2000}, so that its optical flux decreases by
25\% in 3 weeks, for a 0.3 mag change.  Such an effect is small
compared to the large dynamic range of the correlation, which covers 8
optical magnitudes and 3 orders of magnitude in X-ray luminosity.  The
X-ray flux of A0620 in outburst was $\sim$50 Crab \citep{esin2000},
which translates to $\sim$8$\times$10$^{37}$ erg s$^{-1}$ at the
appropriate distance (1.05$\pm$0.40 kpc; \citealp{shahbaz1994a0620}).
Applying the 75 day optical decay time to the peak optical magnitude
($V=11.2$; \citealp{liu2001}), A0620 was $V=11.5$ three weeks after
peak.  Assuming an extinction of $A_V=1.2$ \citep{liu2001},
M$_V$=0.2$\pm$0.9 at that time.  The orbital period prediction from
these values is 0.8$^{+7.6}_{-0.7}$ days, which is consistent with the
known period of 0.3 days \citep{liu2001}.  Similar results are seen
when the correlation is applied to other classical systems, including,
for example, 4U~1543-47 \citep{williams2005bh1}.  The predicted period
range is therefore reliable for classical XRNe even if the optical
observation is 3 weeks after the X-ray observation.

In addition, the correlation even holds for several more recent X-ray
transient sources that have exhibited complex light curves, such as
GRO~J1655-40 and XTE~1550-564 \citep{williams2005bh1}.  We tested the
effects of the 3 week interval between the X-ray and optical
observations of r2-71 on our period prediction using the complicated
optical and X-ray lightcurves of the recent XRN
XTE~J1550-564. Inspection of the lightcurves of \citet{jain2001}
suggests that if we observed XTE~J1550-564 $\sim$8 days after its
X-ray peak, when its X-ray luminosity was $\sim$4$\times$10$^{37}$ erg
s$^{-1}$ (for a distance of 5.3$\pm$2.3 kpc;
\citealp{orosz2002j1550}), and then observed the location in the
optical 3 weeks later, we would have seen the counterpart at
$V\sim19$.  Applying an extinction of $A_V=4.75$
\citep{orosz2002j1550} implies M$_V = 0.6^{+1.2}_{-0.8}$.  Putting
these numbers into the empirical correlation provides a period
prediction of $0.7^{+4.8}_{-0.6}$ days.  If we were fortunate enough
to catch the brightest X-ray flare, with a flux of
1.6$\times$10$^{-7}$ erg cm$^{-2}$ s$^{-1}$ \citep{sobczak2000}, the
source luminosity would have been L$_X\sim5\times$10$^{38}$ erg
s$^{-1}$.  In this case, we would have measured $V\sim17.5$ three
weeks later \citep{jain1999}, 0.9 mag fainter than the peak of
$V=16.6$ \citep{liu2001}.  The later optical measurement would have
yielded M$_V = -0.9^{+1.2}_{-0.8}$ and a period prediction of
$1.1^{+13.3}_{-1.0}$ days.  The actual period of XTE~J1550-564 is 1.55
days \citep{orosz2002j1550}, within the predicted range.

Succinctly, the {\it HST} data show that r2-71 is not an HMXB, and
therefore may be an LMXB.  The \citet{vanparadijs1994} correlation
provides reliable orbital period range predictions for such objects
even when the observations are separated by 3 weeks and the relation
is applied to a complex transient lightcurve.  We therefore apply the
correlation to our measurements of r2-71 under the assumption that, as
in the above Galactic examples, the errors in absolute $V$ magnitude
and in the correlation are sufficient to account for complications in
the lightcurve and the 3-week gap between X-ray and optical
observations. We note that these predictions rely on the assumption
that r2-71 behaves in a similar way to many Galactic XRNe.  

Our $B$-band brightness upper-limit of $B\geq25.5$, from our
completeness results, can be converted to a $V$-band luminosity by
assuming the same foreground extinction we assumed for the X-ray
spectral fit and converting to optical extinction using the relation
of \citet{predehl1995}.  Assuming $m-M = 24.47$, M$_B \geq 0.6$.  Then
using the mean $B-V$ colors of Galactic LMXBs in the \citet{liu2001}
catalog (-0.09 +/- 0.14), M$_V \geq 0.5$. Placing this upper-limit on
the optical luminosity and our 0.3--7 keV X-ray luminosity of 3.9
$\times 10^{37}$ erg s$^{-1}$ into the \citet{vanparadijs1994}
correlation, including their quoted errors, we obtain a prediction for
the period of the LMXB system r2-71 of $P\leq1.6$ days.

\section{Conclusions}

We have constrained the X-ray and optical properties of a repeating
X-ray transient source in the M31 bulge, which we have named
CXOM31~J004233.4+411703 or r2-71.  This source has undergone at least
two X-ray outbursts brighter than 10$^{37}$ erg s$^{-1}$ in the past
two decades.  Previous X-ray observations reveal that the source has
varied by at least a factor of 100 in X-ray luminosity, and our {\it
Chandra} monitoring program shows that the outburst in May of 2004 had
an $e$-folding decay time of less than a month.  The observed activity
of the source from 1979 to the present suggests that it has a duty
cycle of 0.04$^{+0.06}_{-0.03}$.

Optical observations of the location of r2-71 with $HST$ ACS before,
during, and after the X-ray outburst show no clear optical counterpart
to this transient X-ray event in the M31 bulge.  The stellar content
of the region rules out the presence of an HMXB transient system at
the location of r2-71.  We therefore assume r2-71 is an LMXB system.
No reliable variability was detected in the r2-71 error circle, so
that we did not detect the optical counterpart of the XRN.  A
difference image of the region confirms the lack of significant
variability.  The optical data therefore place an upper-limit on the
$B$-band brightness of the outburst of $B\geq25.5$.  The corresponding
upper-limit on the $V$-band luminosity along with the X-ray luminosity
measured from the {\it Chandra} spectrum provide a prediction of
$\leq$1.6 days for the orbital period of the LMXB system.

Support for this work was provided by NASA through grant number
GO-9087 from the Space Telescope Science Institute and through grant
number GO-3103X from the {\it Chandra} X-Ray Center.  MRG acknowledges
support from NASA LTSA grant NAG5-10889.  SSM acknowledges the support
of the HRC contract NAS8-03060.  JEM acknowledges the support of NASA
grant NNG0-5GB31G.

%\bibliography{apjmnemonic,references}
%\bibliographystyle{apj}

\end{document}